\documentclass[aps,pra,reprint,superscriptaddress,amsmath,amssymb,showpacs]{revtex4-1}
\usepackage[pdftex]{graphicx}% Include figure files
\usepackage{dcolumn}% Align table columns on decimal point
\usepackage{bm}% bold math
\usepackage{epsfig}
\usepackage{amsmath}
\usepackage{amsfonts}
\usepackage{subfigure}
\usepackage{float}

\newcommand{\ba}{\begin{eqnarray}}
\newcommand{\ea}{\end{eqnarray}}
\newcommand{\be}{\begin{equation}}
\newcommand{\ee}{\end{equation}}
\newcommand{\bea}{\begin{eqnarray}}
\newcommand{\eea}{\end{eqnarray}}

\usepackage{theorem}

\theorembodyfont{\rmfamily}

\theoremstyle{break}

\def\QED{~\rule[-1pt]{5pt}{5pt}\par\medskip}

\def\Z {\mathbb{Z}}

\def\a{\mathfrak{a}}

\newcommand{\fk}{\mathfrak{k}}
\newcommand{\fp}{\mathfrak{p}}
\newcommand{\fa}{\mathfrak{a}}

\newcommand{\umax}{u_{\mathrm{max}}}
\newcommand{\vmax}{v_{\mathrm{max}}}
\newcommand{\tmin}{T_{\mathrm{min}}}
\newcommand{\ket}[1]{\ensuremath{| #1 \rangle}{}}
\newcommand{\SU}{\mathrm{SU}}

\usepackage{xcolor}

\usepackage[bookmarks=false,pdfstartview={FitH}]{hyperref}

\allowdisplaybreaks

\begin{document}

%\preprint{}

\title{Time-optimal polarization transfer from an electron spin to a nuclear spin}


%\author{}
%\thanks{}
%\affiliation{}

\author{Haidong Yuan}
\email[]{hdyuan@mae.cuhk.edu.hk}
%\thanks{}
\affiliation{Department of Mechanical and Automation Engineering, 
The Chinese University of Hong Kong, Shatin, Hong Kong}

\author{Robert Zeier}
\email[]{robert.zeier@ch.tum.de}
\affiliation{%
Department Chemie,
Technische Universit{\"a}t M{\"u}nchen,
Lichtenbergstrasse 4,
85747 Garching, Germany}%

\author{Nikolas Pomplun}
\email[]{nikolas.pomplun@bruker.com}
\affiliation{%
Department Chemie,
Technische Universit{\"a}t M{\"u}nchen,
Lichtenbergstrasse 4,
85747 Garching, Germany}%
\affiliation{Bruker BioSpin GmbH, Silberstreifen 4, 76287 Rheinstetten, Germany}

\author{Steffen J. Glaser}
\email[]{steffen.glaser@tum.de}
%\thanks{}
\affiliation{%
Department Chemie,
Technische Universit{\"a}t M{\"u}nchen,
Lichtenbergstrasse 4,
85747 Garching, Germany}%

\author{Navin Khaneja}%
%\email[]{}
\affiliation{%
Department of Electrical Engineering, 
IIT Bombay, Powai, Mumbai 400 076, India}%


%\date{\today}
\date{September 7, 2015}

\begin{abstract}
Polarization transfers from an electron spin to a nuclear spin are 
essential for various physical tasks, such as dynamic nuclear 
polarization in nuclear magnetic resonance and quantum state 
transformations on hybrid electron-nuclear spin systems.
We present time-optimal schemes
for electron-nuclear polarization transfers which improve on conventional 
approaches and will have wide applications.
\end{abstract}

%03.67.Ac	Quantum algorithms, protocols, and simulations
%33.25.+k	Nuclear resonance and relaxation
%33.35.+r	        Electron resonance and relaxation
%02.30.Yy	Control theory
%\keywords{}

\maketitle

\section{Introduction}
As the gyromagnetic ratio of an electron is two to three orders 
of magnitude larger than the one of a nucleus, electron spins 
are much easier polarized than nuclear spins. This offers a 
way to improve the polarization of nuclear spins by transferring 
polarization from electron spins to nuclear spins; much higher 
nuclear spin polarization can be achieved as compared to a direct 
polarization. This idea has been widely used in various physical 
settings, for example dynamic nuclear polarization (DNP) 
\cite{SJ:2001,BFH87,WeisGriffin06,MTAPZB07,MDBH+08,GBVLPHEGDP12} 
employs this idea to dramatically improve the sensitivity of 
nuclear magnetic resonance (NMR) \cite{Levitt08,EBW87}. It is 
also frequently used on various hybrid electron-nuclear spin 
systems, such as organic single crystals \cite{MMS:2003}, 
endohedral fullerenes \cite{SM:2008,NMHM:2008,MTAPLB:2007}, 
phosphorous donors in silicon crystals \cite{MTBSLASHAL:2008}, 
and nitrogen-vacancy centers in diamond
\cite{JGPDGW:2004,CGTZJWHL:2007,WWZJ+14,MKCM+14}. 
For example, in the case of nitrogen-vacancy centers in diamond,
efficient polarization transfers are used to initialize the quantum
state of nuclear spins for quantum information processing.

Efficient polarization transfers are practically achieved by 
properly engineered pulse sequences whose design is studied
in the field of quantum control 
\cite{WRD93,RiceZhao00,ShapiroBrumer03,Tannor07,dAll08,BCR10}.
In recent years, significant progress has been made in quantum control for 
both numerical 
\cite{BryHo75,CNM86,Rabitz87,SWR88,Krotov96,OZR99,KK99,PK03,OTR04,GRAPE,
TVKKGN09,MSGFGSS11,FSGK11,EMT11} and analytical 
\cite{Pont62,Jurdjevic97,BonnardChyba03,BoscainPiccoli04} methods.
Extensive knowledge has been gained on optimal pulse sequences for two- 
and three-level systems
\cite{TB99,BCGGJ02,BCC05,BM06,SKJ07,BS09,BCS09,LZBGS10,LZGS11,ZLSBG11,Boozer12,
MCA13,GGS13,LAGS13,ADA15}, two uncoupled spins \cite{ALZBGS10,AAPPTZGS12},
and two coupled spins \cite{Khaneja01b,BCL+02,VHC02,HVC02,YK05,RKG:2002,KKG:2005}.
Further advances have been made on how to optimally control 
multiple coupled spins
\cite{KGB02,KG02,RKG03,Bose03,ZGB04,SKG04,SKG05,SHSKG05,CHKO06,KHSY07,YGK07,YZK08,
SP09,BP09,WBBS10,WBSB10,CHKO11,YK11,LSS12,CK12,NZN12,CK13,BCS13,BCPS14,YWZGK14,DZGS14}. 
These methods have been successfully applied in NMR \cite{FPBT02,NKGK10} 
to designing broad-band \cite{SRLKG03,SRLKG04,KSKGL04} and
decoupling pulse sequences \cite{PrySen08,NHKG09,SG12,ZSGH13,KMSH13,SWGSSG14}.
They have also been utilized in magnetic resonance imaging
\cite{CNM86,GXKF09,LM11,LZJGS12} and 
electron paramagnetic resonance \cite{SZEGSGP12}.

In this article, we consider time-optimal pulse sequences for 
polarization transfers from an electron spin to a nuclear spin. 
Relaxation and decoherence are in practice inevitable and result 
in a loss of signal. But their effect can be 
mitigated by short pulse sequences which
allow for highly sensitive experiments.
We analyze and explain how the form of time-optimal sequences depends 
on the direction of the polarization by studying time-optimal transfers 
for different directions.

Recent analytical \cite{Kha:2007} and numerical \cite{HYRC08,MaxTosNie08} studies
focused on low-field single-crystal experiments, where the nuclear Larmor frequency and
\emph{pseudo-secular} hyperfine interaction (see Sect.~3.5 of Ref.~\cite{SJ:2001})
are comparable in magnitude.
As in \cite{ZYK:2008,pomplun:2008,pomplun_thesis,pomplun:2010},  we focus here on 
the cases of
\emph{secular}   hyperfine coupling (see Sect.~3.5 of Ref.~\cite{SJ:2001}).
These assumptions are satisfied in liquid-state and high-field solid-state DNP.

We analyze two particular cases of polarization transfers and
determine the corresponding time-optimal sequences. 
In Section~\ref{sec:theory}, we consider the transfer from 
the state $S_z$ of the electron spin to the state $I_z$
of the nuclear spin. The second time-optimal transfer from $S_z$ to $I_x$
is presented in Section~\ref{sec:theory2}. And most interestingly, the 
corresponding optimal transfer time is shorter by $78.5\%$ when compared to 
the transfer from $S_z$ to $I_z$, which highlights that the transfer efficiency depends
crucially on the target state of the nuclear spin.
We discuss our 
results in Section~\ref{sec:disc}, and the possibility of a non-sinusoidal carrier wave form
is entertained in Section~\ref{gen}. We conclude in Section~\ref{conc}, and
certain details are relegated to Appendices~\ref{app_A} and \ref{app:surj}.

\section{Transfer from $S_z$ to $I_z$\label{sec:theory}}

In this section, we study the polarization transfer from
the initial state $S_z$  to the final state $I_z$ \footnote{Note that 
we only consider the
(rescaled) traceless part $\rho$ of the actual density matrix $b \rho+\frac{1}{4}I$,
where $b$ denotes the Boltzmann factor. As the identity component does not evolve,
it can be neglected.}. 
We assume a secular hyperfine coupling (see Sect.~3.5 of 
Ref.~\cite{SJ:2001}).
In the lab frame, the resulting Hamiltonian is given by
\begin{equation} \label{lab_frame}
H=\omega_S S_z+\omega_II_z+2\pi A S_zI_z+2\pi \tilde{u}_x(t) S_x+ 
2\pi \tilde{v}_x(t) I_x,
\end{equation}
where $\omega_S$ and $\omega_I$ denote
the respective Larmor frequencies of the electron 
and the nuclear spin, $A$ represents the strength of the secular hyperfine
coupling, $\tilde{u}_x(t)$ and $\tilde{v}_x(t)$ are the amplitudes of
the control fields.
Here,  $S_{j}=(\sigma_{j}\otimes\sigma_{0})/2$ acts on the electron spin
and $I_{k}= (\sigma_{0}\otimes\sigma_{k})/2$  acts on the nuclear spin
with $j,k \in \{x,y,z\}$, where
$
\sigma_{0}:=
\left(
\begin{smallmatrix}
1 & 0 \\
0 & 1
\end{smallmatrix}
\right)
$ denotes the identity matrix and
the Pauli matrices are 
$\sigma_{x}:=
\left(\begin{smallmatrix}
0 & 1 \\
1 & 0
\end{smallmatrix}
\right)$,
$
\sigma_{y}:=
\left(
\begin{smallmatrix}
0 & -i \\
i & \phantom{-}0
\end{smallmatrix}
\right)
$, and
$
\sigma_{z}:=
\left(
\begin{smallmatrix}
1 & \phantom{-}0 \\
0 & -1
\end{smallmatrix}
\right)
$.
For typical NMR settings, only
a single radio-frequency coil is used
which can be assumed to be oriented along the $x$ axis of the lab frame.
Hence, only a single control $\tilde{v}_x(t)$  appears
for the nuclear spin in the lab frame Hamiltonian of Eq.~\eqref{lab_frame}.
We assume in this work that the carrier wave form for the
nuclear spin
has a sinusoidal shape, i.e.
$\tilde{v}_x(t)=
v(t) \cos[\omega_I^{\mathrm{rf}} t + \phi(t)]$ with amplitude $v(t)\leq 2 \vmax$ and
phase $\phi_0$. Here, $\omega_I^{\mathrm{rf}}$ is the carrier frequency of the
radio-frequency irradiation and $2 \vmax$ denotes the maximal control amplitude
(in the lab frame).
This choice of $\tilde{v}_x(t)$ is motivated by the properties 
(e.g., bandwidth limitations) of the usually
available wave form generators and
amplifiers. 
More general carrier wave forms are discussed in Section~\ref{gen}.

By switching to the
rotating frame of $ \omega_S S_z+ \omega_I^{\mathrm{rf}} I_z$
corresponding to the carrier frequencies $\omega_S $ and
$\omega_I^{\mathrm{rf}}=\omega_I-\omega_I^{\mathrm{off}}$
and applying the rotating wave approximation, we get an effective Hamiltonian
\begin{align}
H_{\mathrm{rot}}=&+\omega_I^{\mathrm{off}} I_z +2\pi A S_zI_z \nonumber \\
&+ H^{\mathrm{mw}}_{\mathrm{rot}}
+ H^{\mathrm{rf}}_{\mathrm{rot}}, \text{ where} \label{Ham} \\
H^{\mathrm{mw}}_{\mathrm{rot}}:=\, &2\pi u_x(t)S_x+2\pi u_y(t)S_y, \nonumber \\
H^{\mathrm{rf}}_{\mathrm{rot}}:=\, &2\pi v_x(t)I_x+2\pi v_y(t)I_y. \nonumber
\end{align}
One can obtain any desired offset term $\omega_I^{\mathrm{off}} I_z$ in the drift term of $H_{\mathrm{rot}}$ in 
Eq.~\eqref{Ham}
by suitably choosing the carrier frequency $\omega_I^{\mathrm{rf}}$
\footnote{In \cite{ZYK:2008}, the choice for $\omega_I^{\mathrm{off}}$ was
$\omega_I^{\mathrm{off}}=\pi A$.}.
For simplicity, $\omega_I^{\mathrm{off}}$ is set to zero
in the following.
The microwave-frequency control pulses on the electron spin and 
the radio-frequency control pulses
on the nuclear spin are given by
$H^{\mathrm{mw}}_{\mathrm{rot}}$ and $H^{\mathrm{rf}}_{\mathrm{rot}}$, respectively.
The control amplitudes $u_x(t)$, $u_y(t)$, $v_x(t)$, and $v_y(t)$ satisfy the bounds
\begin{equation*}
\sqrt{u_x^2(t)+u_y^2(t)}\leq \umax\; \text{ and }\; \sqrt{v_x^2(t)+v_y^2(t)}\leq \vmax,
\end{equation*}
where $\umax$ and $\vmax$ denote the maximal available amplitudes of the control fields  
in the rotating frame for a given experiment.
This is a result of the rotating wave approximation, which reduces the maximal 
control amplitude of $2\vmax$ in lab frame to $\vmax$ in the rotating frame \cite{EBW87}.
In the following, we will assume that $\umax \gg A \gg \vmax$ and 
neglect the time needed to apply operations that can be generated by 
the hyperfine coupling and the controls on the electron spin
\footnote{Strictly speaking, our results
are also applicable when only the conditions
$A \gg \vmax$ and $\umax \gg \vmax$ 
are fulfilled.}.

Time-optimal transformations are essentially only limited by the
weak controls on the nuclear spin. The optimal strategy to achieve a desired 
transfer can be inferred from the structure of cosets with respect to the fast 
operations \cite{Khaneja01b,KGB02,ZYK:2008}. Here, the fast operations are given 
by the hyperfine coupling and the strong controls on the electron spin. The 
transformation $U$ which transfers $S_z$ to $I_z=U S_z U^{-1}$ will be suitably 
decomposed into a product $U=U_2 U_1$. The unitary $U_1$ transfers the initial 
state $S_z$ to the intermediate state $2 S_z I_z=U_1 S_z U_1^{-1}$, and it 
can be generated using only fast operations. In addition, the unitary $U_2$ 
transfers the intermediate state $2 S_z I_z$
to the final state $I_z=U_2 (2 S_z I_z) U_2^{-1}$,
and one has to use the weak controls on the nuclear spin in order to generate $U_2$.
Below, we will provide a time-optimal scheme to produce $U_2$. This results also in
a time-optimal scheme for $U$ as any faster scheme for $U$ would also
imply a faster one for $U_2=U U_1^{-1}$.

\begin{figure}[t]
\includegraphics{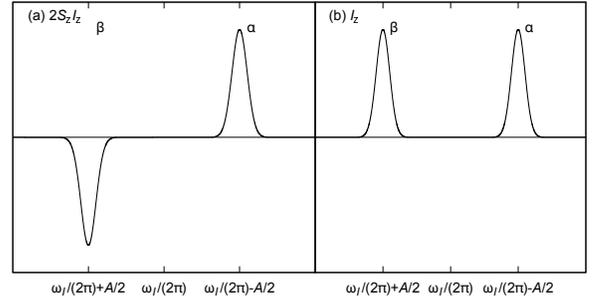}
\caption{Schematic depiction of the absorption profiles for 
(a)~$2 S_z I_z$ and (b) $I_z$. \label{fig:delta_one}}
\end{figure}

\begin{figure}[t]
\includegraphics{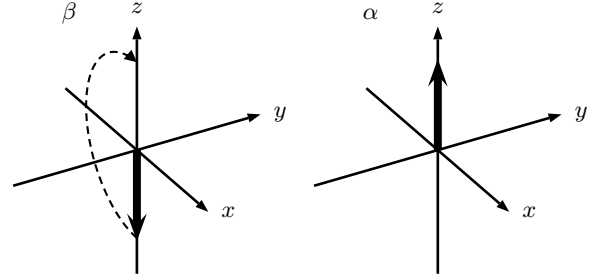}
\caption{In the polarization transfer from $2S_zI_z$ to $I_z$, the $\beta$ component
of the nuclear spin doublet is rotated by 
an angle of $\pi$ around the $y$ axis and the $\alpha$ component is 
left invariant (both visualized in the interaction frame). Note that the electron spin is 
in the state $\ket{\beta}$ on the left-hand side and in the state
$\ket{\alpha}$ on the right-hand side. \label{fig:one}}
\end{figure}

The polarization transfer from $S_z$  to $I_z$
can be decomposed into the following steps \footnote{The notation 
$A \xrightarrow{B} C$ describes
that polarization is transferred from $A$ to $C$ by applying the unitary 
transfer $C=\exp(-i B) A \exp(i B)$.}:  
\begin{equation}\label{eq:transfer}
S_z \xrightarrow{\tfrac{\pi}{2}S_y}
S_x\xrightarrow{\pi S_zI_z}2S_yI_z\xrightarrow{\tfrac{\pi}{2}S_x}2S_zI_z
\xrightarrow{\pi S^{\beta} I_y } I_z,
\end{equation}
where we denote $$S^{\alpha}:=\left(
\begin{smallmatrix}
1 & 0 \\
0 & 0
\end{smallmatrix}
\right)
\otimes\sigma_{0}\,\text{ and }\,S^{\beta}:=\left(
\begin{smallmatrix}
0 & 0 \\
0 & 1
\end{smallmatrix}
\right)
\otimes\sigma_{0},$$ then $\pi S^{\beta} I_y =-\pi S_zI_y+\pi I_y/2$.
As shown in Eq.~\eqref{eq:transfer}, the polarization transfer from 
$S_z$ to $2S_zI_z$ is accomplished using an INEPT-type 
transfer \cite{MF79,EBW87}:
First, we apply a hard ${\pi}/{2}$ pulse to the electron spin along the 
$+y$ direction (i.e.\ $S_y$).
Then, we let the hyperfine coupling evolve for the duration of $1/(2A)$ 
units of time. Another hard ${\pi}/{2}$ pulse on the electron spin along the $+x$ 
direction completes the transfer to $2S_zI_z$. All of these steps take negligible 
time, since they are either local operations on the electron spin or 
operations which can be generated by the coupling. 
In conclusion, we can completely focus on the last step in Eq.~\eqref{eq:transfer}
where we need to generate the propagator
\begin{equation}
U^\beta_y(\theta)=\exp(-i\theta S^\beta I_y)=\exp[-i(-\theta S_zI_y+
\tfrac{\theta}{2}I_y)] \label{eq:expone}
\end{equation}
for $\theta=\pi$. 
The operator in the exponent of $U^\beta_y(\theta)$ in Eq.~\eqref{eq:expone} 
is a single-transition operator \cite{EBW87,SJ:2001}. In particular, the operator 
$U^\beta_y(\pi)=\exp(-i \pi S^\beta I_y)$ describes a transition-selective $\pi$
rotation around the $y$ axis in the subspace spanned by the basis states 
$\ket{\beta \alpha}$ and $\ket{ \beta \beta}$, where the subspace corresponds to 
the $\beta$ component of the nuclear spin doublet at frequency 
$\omega_I/(2 \pi) + A/2$ \cite{Keeler2010} as shown in 
Figures~\ref{fig:delta_one} and \ref{fig:one}. Here,
$\ket{\alpha}$ and $\ket{\beta}$ are eigenstates of $S_z$
and $I_z$, e.g., $S_z \ket{\alpha} = \ket{\alpha} /2$  and $S_z \ket{\beta} = - \ket{\beta} /2$.

In the following, we determine a time-optimal scheme to produce
the unitary $U^\beta_y(\pi)$.
The set of all unitaries which transfer $2S_zI_z$ to $I_z$ are
discussed in Appendix~\ref{appendix_Iz}, where we also show 
by extending the results in the current section
that choosing a different element from this set of unitaries
does not lead to a shorter transfer time.

To determine the optimal transfer, we switch to the interacting frame 
of $2\pi A S_zI_z$ by applying the transformation
$\exp(i2\pi A S_z I_z t) H_{\mathrm{rot}} \exp(-i2\pi A S_z I_z t)$. 
The Hamiltonian of Eq.~\eqref{Ham}
changes to \footnote{In the interaction frame, $H_{\mathrm{rot}}$ 
is transformed
into
$H_{\mathrm{int}}=\exp(i2\pi A S_z I_z t) H_{\mathrm{rot}} \exp(-i2\pi A S_z I_z t)
- 2\pi A S_z I_z$.}
\begin{align}
H_{\mathrm{int}} 
=&+2\pi u_x(t)[\cos({\pi At})S_x-\sin({\pi At})2S_yI_z] \nonumber \\
&+2\pi u_y(t)[\cos({\pi At})S_y+\sin({\pi At})2S_xI_z] \nonumber \\
&+2\pi v_x(t)[\cos({\pi At})I_x-\sin({\pi At})2S_zI_y] \nonumber \\
&+2\pi v_y(t)[\cos(\pi At)I_y+\sin({\pi At})2S_zI_x] \label{H_transformed} \\
\intertext{which can be also written as} 
H_{\mathrm{int}} =&+2\pi u_x(t)[\cos({\pi At})S_x-\sin({\pi At})2S_yI_z] \nonumber \\
&+2\pi u_y(t)[\cos({\pi At})S_y+\sin({\pi At})2S_xI_z] \nonumber \\
&+2\pi [v_x(t)\cos({\pi At})I_x+v_y(t)\sin({\pi At})2S_zI_x] \nonumber \\
&-2\pi [v_x(t)\sin({\pi At})-v_y(t)\cos(\pi At)]
S^\alpha I_y 
\nonumber \\
&+2\pi [v_x(t)\sin({\pi At})+v_y(t)\cos(\pi At)]
S^\beta I_y,
\label{Eq_complex}
\end{align}
where $S^\alpha I_y =S_zI_y+I_y/2$ and  $S^\beta I_y=-S_zI_y+I_y/2$.

We aim at generating the operator $U_y^\beta(\pi)$ in minimum time, which corresponds 
to maximizing the coefficient $2\pi [v_x(t)\sin({\pi At})+v_y(t)\cos(\pi At)]$
in front of $S^\beta I_y$. The Cauchy-Schwarz inequality implies
\begin{align*}
&[v_x(t)\sin({\pi At})+v_y(t)\cos(\pi At)]^2 \leq \\
&[v^2_x(t)+v^2_y(t)][\sin^2({\pi At})+\cos({\pi At})^2]
\leq \vmax^2,
\end{align*}
where the second inequality is a consequence of the constraints 
on the amplitude of the control fields. 
The maximal value of $2\pi [v_x(t)\sin({\pi At})+v_y(t)\cos(\pi At)]$ is 
denoted by $2\pi \vmax$ and it can be achieved by choosing the controls
$u_x(t)=u_y(t)=0$, $v_x(t)=\vmax \sin(\pi At)$, $v_y(t)=\vmax \cos(\pi At)$.
To understand that this choice generates the desired operator,
one can substitute the controls in the Hamiltonian with the chosen
values and obtains
\begin{align*}
H_{\mathrm{int}}=&+2\pi \vmax \sin({\pi At}) \cos({\pi At})I_x\\
&+2\pi \vmax\cos({\pi At}) \sin({\pi At})2S_zI_x \\
&+2\pi \vmax \cos(2\pi At) S^\alpha I_y+2\pi\vmax S^\beta I_y.
\end{align*}
Since $A \gg \vmax$, average Hamiltonian theory implies that
the first three terms average out to zero; 
and one is left with the desired Hamiltonian $2\pi\vmax S^\beta I_y$.

The minimum time to generate $U^\beta_y(\pi)$ is then fixed by 
the relation $2\pi \vmax \tmin = \pi$, and one obtains $$\tmin=1/(2\vmax).$$
The presented time-optimal control corresponds to a radio-frequency irradiation at frequency
$\omega_I/(2 \pi) + A/2$ with duration $\tmin$, which results in a 
transition-selective inversion
of the $\beta$ line of the nuclear spin doublet. This belongs to the class of 
controls presented in 
Ref.~\cite{ZYK:2008} and is also closely related to selective population inversion (SPI) 
experiments \cite{FP73,PW73,CHM80,Sor89}.

\begin{figure}[b]
\includegraphics{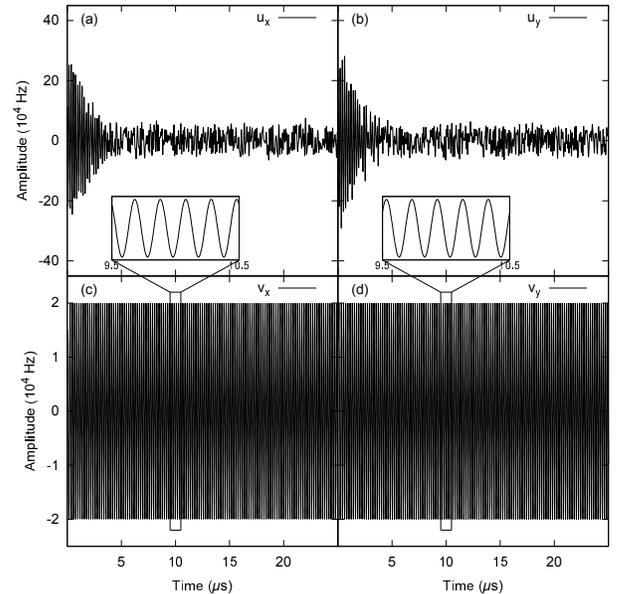}
\caption{Numerically optimized pulses for the polarization transfer from $S_z$ to $I_z$. 
The coupling strength is 10 MHz and the bounds on the micro-wave and  
radio-frequency amplitudes 
are  $\umax=1 \text{ MHz}$ 
and $\vmax=20 \text{ kHz}$, respectively. The maximal transfer efficiency is reached 
after $25 \mu s$.
The insets show magnified parts of the controls $v_x(t)$ and $v_y(t)$ 
in order to illustrate their form.
\label{fig:vx1}}
\end{figure}

We can also compute the maximal transfer efficiency $\eta_{max}(T)$  
for a given time $T$.
The operator $U^\beta_y(\theta)$ transfers the state $2S_zI_z$ to the state
\begin{align}
&U^\beta_y(\theta)(2S_zI_z)[U^\beta_y(\theta)]^{\dagger}=\nonumber \\
&\cos^2(\tfrac{\theta}{2})2S_zI_z+\cos(\tfrac{\theta}{2})\sin(\tfrac{\theta}{2})2S_zI_x 
\nonumber \\
&-\cos(\tfrac{\theta}{2})\sin(\tfrac{\theta}{2})I_x+\sin^2(\tfrac{\theta}{2})I_z. 
\label{eq:top}
\end{align}
For $\theta=2\pi \vmax T$, we get
the maximal transfer efficiency 
\begin{equation}\label{eq:transfer_fct_one}
\eta_{max}(T)=\sin^2(\tfrac{\theta}{2})=\sin^2(\pi \vmax T)
\end{equation}
for the transfer to $I_z$. Note that $\eta_{max}(\tmin)=1$.

We compare our analytic results with numerical optimizations for
achieving the transfer from $S_z$ to $I_z$ as shown in Fig.~\ref{fig:vx1},
cf.\ \cite{pomplun:2008,pomplun_thesis,pomplun:2010}.
For these optimizations, the hyperfine coupling constant is chosen 
as $A=10 \text{ MHz}$ and the maximal allowed radiation amplitudes are set to 
$\umax=1 \text{ MHz}$ and $\vmax=20 \text{ kHz}$ \footnote{Even though we have 
$\umax < A$, the crucial requirements $A \gg \vmax$ and $\umax \gg \vmax$ 
for optimality are fulfilled.}. In Fig.~\ref{fig:vx1}, the transfer is completed 
after $25 \mu s = 1/(2\vmax)$ units of time which agrees with the analytically 
computed time. Moreover, the form of the numerically optimized controls compares 
nicely with the analytic results: the values of $u_x(t)$ and $u_y(t)$ are most of 
the time small (except for the beginning), and $v_x(t)$ and $v_y(t)$ have a 
sinusoidal form with the maximal allowed amplitude.

\section{Transfer from $S_z$ to $I_x$ or $I_y$\label{sec:theory2}}

We analyze now how to time-optimally transfer polarization from 
the state $S_z$ to $I_x$ (and similarly for the 
transfer to $I_y$).
The considered transfer consists of the following steps:
$$S_z\xrightarrow{\tfrac{\pi}{2}S_y}
S_x\xrightarrow{\pi S_zI_z}2S_yI_z\xrightarrow{\tfrac{\pi}{2}S_x}
2S_zI_z\xrightarrow{\tfrac{\pi}{2} S^{\alpha}I_y-\tfrac{\pi}{2} 
S^{\beta}I_y} I_x,$$
where $\tfrac{\pi}{2} S^{\alpha}I_y-\tfrac{\pi}{2} S^{\beta}I_y=\pi S_zI_y$.
As in Sec.~\ref{sec:theory}, we can focus on generating the 
final propagator $\tilde{U}_2=\exp(-i\pi S_zI_y)$ in the product
$\tilde{U}=\tilde{U}_2 U_1$. The transfer $I_x=\tilde{U} S_z \tilde{U}^{-1}$
is decomposed into a fast transfer to the 
intermediate state $2 S_z I_z=U_1 S_z U_1^{-1}$ and a slow transfer 
to final state $I_x=\tilde{U}_2 (2 S_z I_z) \tilde{U}_2^{-1}$.
Building on the results in this section, we prove in Appendix~\ref{appendix_Ix}
that one cannot reduce the transfer time by substituting $\tilde{U}_2$
with a different unitary $V$ satisfying $I_x=V (2 S_z I_z) V^{-1}$.

\begin{figure}[t]
\includegraphics{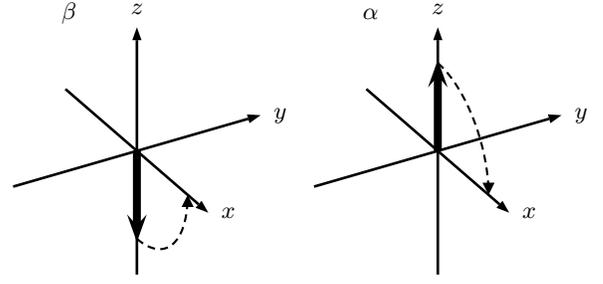}
\caption{In the polarization transfer from $2S_zI_z$ to $I_x$ the $\beta$ component
of the nuclear spin doublet is rotated by $-\pi/2$ around the $y$ axis and the 
$\alpha$ component
is rotated by $\pi/2$ around the $y$ axis (both visualized in the interaction frame).
\label{fig:two}}
\end{figure}

Previously, the propagator $\exp(-i\pi S_zI_y)$ in the final step 
has been achieved \cite{ZYK:2008}
by applying a transition-selective radio-frequency 
$-\pi/2$ pulse along the $y$ direction at the $\beta$ transition with
frequency $\omega_I/(2 \pi) + A/2$ as well as
a transition-selective radio-frequency 
$\pi/2$ pulse along the $y$ direction at the $\alpha$ transition 
with frequency $\omega_I/(2 \pi) - A/2$, see Fig.~\ref{fig:two}.
In the rotating frame of Eq.~\eqref{Ham}, this irradiation scheme
on the nuclear spin corresponds to a
radio-frequency Hamiltonian of the form 
\begin{align*}
H^{\mathrm{rf}}_{\mathrm{rot}}=&
-2\pi \frac{\vmax}{2} \cos(\pi A t) I_x-2\pi \frac{\vmax}{2} \sin(\pi A t) I_y\\
&+
2\pi \frac{\vmax}{2} \cos(-\pi A t) I_x+2\pi \frac{\vmax}{2} \sin(-\pi A t) I_y\\
=&
-2\pi \vmax \sin(\pi A t) I_y.
\label{Hama}
\end{align*}
Note that in this scheme 
the $\alpha$ 
and $\beta$ transitions can only be irradiated with a radio-frequency amplitude of $\vmax/2$ 
in order not to exceed the maximal available radio-frequency 
amplitude $\vmax$ for the overall irradiation at the nuclear spin.
Hence, the duration for the simultaneous  $\pm\pi/2$ pulses
along the $y$ direction at the $\alpha$ and $\beta$ 
transitions is equal to $1/(2\vmax)$.
This conventional transfer is optimal if one considers
\emph{only} pulses at the frequencies 
$\omega_I/(2 \pi) \pm A/2$  of the nuclear-spin doublet 
(refer to \cite{ZYK:2008} and the discussion in 
Section~\ref{sec:disc}). 

Here, we show that shorter pulses are possible if one considers 
more general transfer schemes. Without exceeding $\vmax$,
shorter pulses can be obtained by 
irradiating at the frequencies $\omega_I/(2 \pi) \pm A/2$
with higher intensity since the resulting higher amplitude 
can be then decreased by irradiating 
at \emph{additional} well selected frequencies.
Our approach is quite effective although it 
might seem counterintuitive at first.

In the interaction frame of $2\pi A S_zI_z$, the Hamiltonian is again given by 
Eq.~\eqref{H_transformed}.
In order to generate the operator $\exp(-i\pi S_zI_y)$ in minimum time, we maximize the 
coefficient $-2\pi v_x(t)\sin(\pi At)$ of $2S_zI_y$. Note that  
\begin{align*}
{-}2\pi v_x(t)\sin(\pi At) &\leq   \\  |{-}2\pi v_x(t)\sin(\pi At)|
& \leq  2\pi \vmax |\sin(\pi At)|,
\end{align*}
where the second equality is implied by the constraint
$|v_x(t)|\leq \sqrt{v_x^2(t)+v^2_y(t)}\leq \vmax$ on the control amplitudes.
Therefore, the maximal value $2\pi \vmax |\sin(\pi At)|$ for
$-2\pi v_x(t)\sin(\pi At)$ can be attained  by choosing the controls 
$u_x(t)=u_y(t)=v_y(t)=0$
and $v_x(t)=-\mathrm{sgn}[\sin(\pi At)]\vmax$. This means that 
$v_x(t)$ is a square wave such that  $v_x(t)=\vmax$ when $\sin(\pi At) < 0$
and $v_x(t)=-\vmax$ when $\sin(\pi At) > 0$. In this case, the radio-frequency Hamiltonian
in the rotating frame is given by
\begin{equation} \label{opt_pulse}
H^{\mathrm{rf}}_{\mathrm{rot}}=
-2\pi \vmax\, \mathrm{sgn}[ \sin(\pi A t)]  \, I_y.
\end{equation}
We obtain the minimum time
$$\tmin=\pi/(8\vmax)$$ for generating $\exp(-i\pi S_zI_y)$ in the interaction frame.
The duration of the transfer is reduced to $78.5\%$ of the length of the conventional
pulse sequence.
By transforming the operator back to the rotating frame, we obtain the operator 
$\exp(-i\phi S_zI_z)\exp(-i\pi S_zI_y)
\exp(i\phi S_zI_z)$ where $\phi=2\pi A \tmin$ denotes the phase accumulated during 
the time $\tmin=\pi/(8\vmax)$.
The effect of this superfluous phase $\phi$ can be reversed using the hyperfine coupling 
$2\pi A S_z I_z$
which takes only a negligible time period as the coupling strength $A$ is much larger than
the control strength $\vmax$ of the nuclear spin.
Thus, the minimum time in the rotating frame is also given by $\pi/(8\vmax)$.
Similarly as in Eq.~\eqref{eq:top}, we compute the maximal transfer efficiency 
\begin{equation}\label{eq:transfer_fct_two}
\eta_{max}(T)=\sin(4\vmax T)
\end{equation}
that can be reached
for the polarization transfer from $S_z$ to $I_x$
in a specified time $T$.

Transferring the state from $S_z$ to $I_y$ is similar. We set 
$u_x(t)=u_y(t)=v_x(t)=0$ 
and maximize the coefficient $-2 \pi v_y(t)\sin(\pi At)$ of  $-2S_zI_x$ in 
Eq.~\eqref{H_transformed} by setting 
$v_y(t)=-\mathrm{sgn}[\sin(\pi At)]\vmax$. 
The minimum time for this case is also given by $\pi/(8\vmax)$.

\begin{figure}[t]
\includegraphics{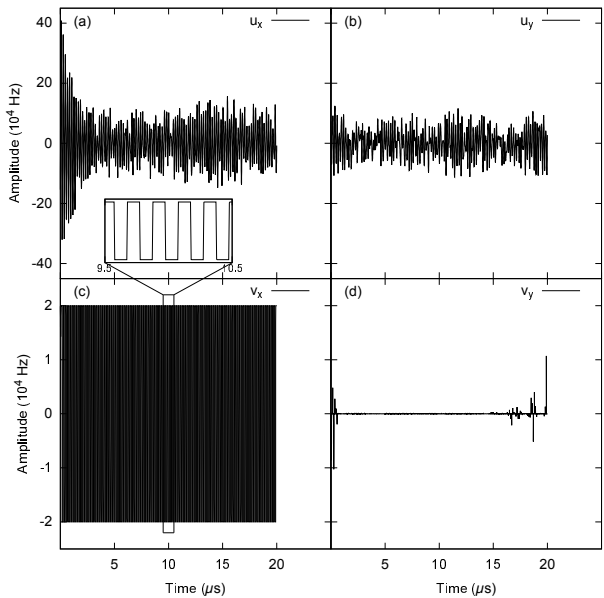}
\caption{Numerically optimized pulses for the polarization transfer from $S_z$ to 
$I_x$. The coupling 
strength is $10 \text{ MHz}$ and the bounds on the  micro-wave  and radio-frequency
amplitudes are given by $\umax=1 \text{ MHz}$ and
$\vmax=20 \text{ kHz}$, respectively. The maximal transfer  efficiency is 
reached after $20 \mu s$.
The inset shows a magnified part of the control $v_x(t)$ 
in order to illustrate its form.
\label{fig:vx2}}
\end{figure}

A numerically optimized pulse sequence for transferring polarization from $S_z$ to $I_x$ is 
shown in Fig.~\ref{fig:vx2}, cf.\ \cite{pomplun:2008,pomplun_thesis,pomplun:2010}. 
The maximal transfer 
efficiency is reached after $20 \mu s$ which is 
consistent with the analytical result  of
$\pi/(8\vmax)\approx 19.635 \mu s$.

\section{Discussion\label{sec:disc}}

\begin{figure}[b]
\includegraphics{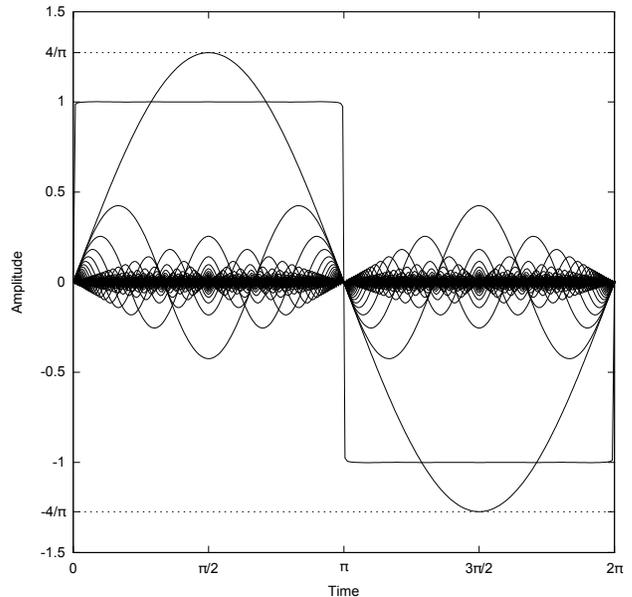}
\caption{Decomposition of a square wave into sine waves: a large number of harmonics 
sum to an
approximate square wave. The first harmonic has an amplitude of $4/\pi$ while the 
square wave has an 
amplitude of $1$.\label{fig:rect}}
\end{figure}

We see that the minimum time for transferring $S_z$ to $I_x$ or $I_y$ is shorter 
by a factor of $\pi/4$ 
when compared to the minimum time for transferring $S_z$ to $I_z$. This factor 
can be explained by 
a closer examination of the pulse sequences. The radio-frequency sequence for the transfer 
from  $S_z$ to $I_z$
shows a sine-cosine wave modulation of maximal amplitude for the $v_x$- and 
$v_y$-components (see Fig.~\ref{fig:vx1}).
However, the radio-frequency sequence 
for the transfer from $S_z$ to $I_x$
consists of a square wave of maximal amplitude for the $v_x$-component of the control 
(see Fig.~\ref{fig:vx2}).
The higher effective amplitude at the two frequencies $\omega_I/(2 \pi) \pm A/2$ and 
the shorter transfer time can be 
explained by  decomposing the square wave into a sum of sine waves:
\begin{equation*}
\label{eq:rect}
f_{\text{square}}(t)=\mathrm{sgn}[\sin(\pi At)]=
\frac{4}{\pi}\sum_{n \text{ odd},\, n\ge 1} \frac{1}{n}\sin(nA t).
\end{equation*}
This is illustrated in Fig.~\ref{fig:rect} where the first sine wave function has 
an amplitude 
which is larger by a factor of $4/\pi$ when compared to the amplitude of the square wave. 
Therefore, the square wave 
contains implicitly a sine wave with an higher effective amplitude.
This implies that the duration of the simultaneous $\pm\pi/2$ rotations of the $\alpha$ and
$\beta$ components of the nuclear spin doublet (see Fig.~\ref{fig:two}) is shorter 
by a factor of $\pi/4$.

\begin{figure}[t]
\includegraphics{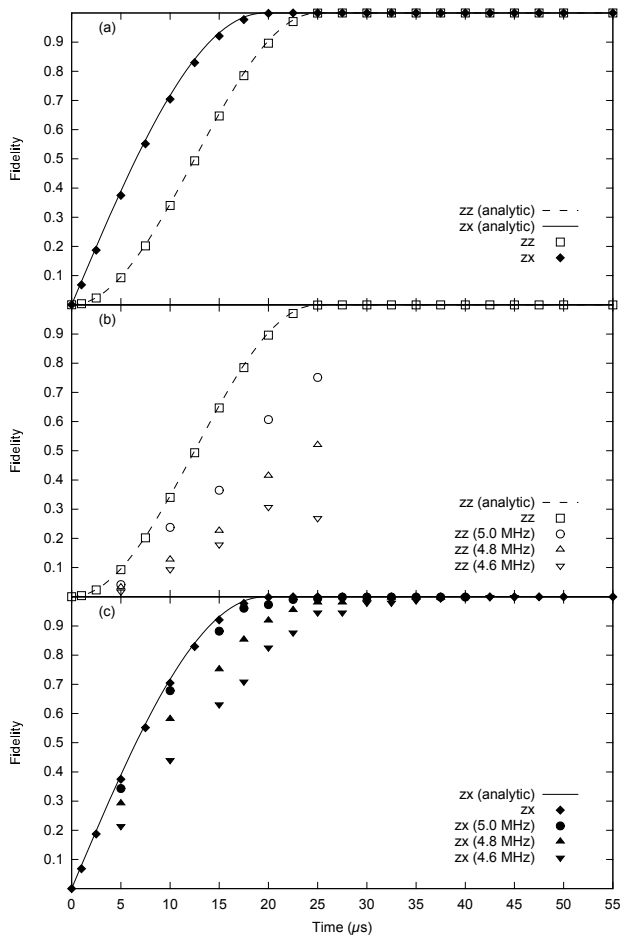}
\caption{The maximal transfer efficiency (i.e.\ fidelity)
is shown for the transfers from $S_z$ 
to $I_z$ and $I_x$ in (a). The coupling strength is 
$10 \text{ MHz}$,
and the control strengths are
$\umax=20 \text{ kHz}$ and $\vmax=1 \text{ MHz}$. 
In (b), data points for the transfer from $S_z$ to $I_z$ 
are shown also for the bandwidth-limited cases
with bounds of $5.0 \text{ MHz}$, $4.8 \text{ MHz}$,  and $4.6 \text{ MHz}$.
Similarly, data points for the transfer from $S_z$ to $I_x$ are shown in (c).}
\label{fig:top}
\end{figure}

The square-modulated transfer sequence is optimal but needs infinite bandwidth. 
We also studied numerically how the
maximal transfer efficiency varies as a function of time and bandwidth limitations. 
The results 
are shown in Fig.~\ref{fig:top}, cf.\ \cite{pomplun:2008,pomplun_thesis,pomplun:2010}. 
In the case of infinite bandwidth, the results are consistent with the analytical results.
The transfer functions $\sin^2(\pi \vmax t)$ and $\sin(4 \vmax t)$  for the 
respective
transfers from $S_z$
to $I_z$ and $I_x$ have been obtained in Eqs.~\eqref{eq:transfer_fct_one} and 
\eqref{eq:transfer_fct_two}.

We compare our results to the time-optimal synthesis of unitary transformations
in \cite{ZYK:2008}. Motivated by energy considerations, only irradiations at the two 
resonance frequencies $\omega_I/(2 \pi) \pm A/2$ of the control system
were considered in \cite{ZYK:2008}. This did not allow for the faster scheme obtained  in 
Section~\ref{sec:theory2} which has been also observed numerically 
in \cite{pomplun:2008,pomplun_thesis,pomplun:2010}. The numerical results in 
Fig.~\ref{fig:top}
also show that the faster scheme of Section~\ref{sec:theory2} is---in a strict 
sense---only applicable in the case of infinite bandwidth. 
It provides in general superior results, but its benefit depends on the 
available bandwidth.

\begin{figure}[t]
\includegraphics[scale=1]{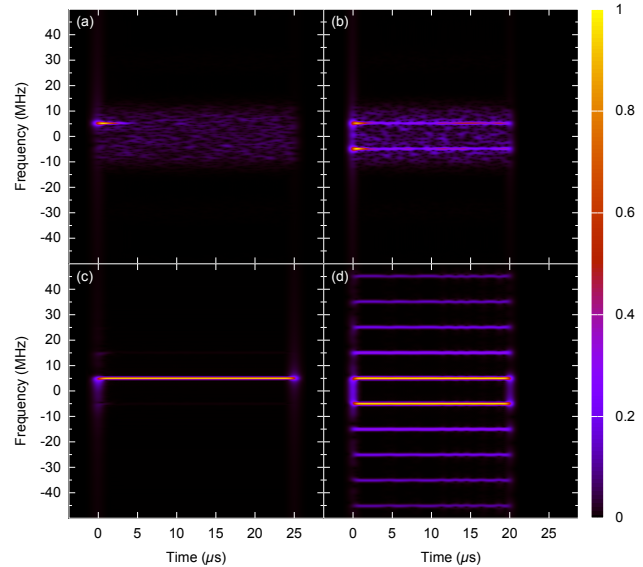}
\caption{(Color online) Normalized amplitudes of the short-time Fourier transform
for different controls: (a) $u$-controls of Fig.~\ref{fig:vx1}, 
(b) $u$-controls of Fig.~\ref{fig:vx2}, (c) $v$-controls of Fig.~\ref{fig:vx1},
and (d) $v$-controls of Fig.~\ref{fig:vx2}.
In (c), (essentially) only the characteristic frequency of $A/2=5 \text{ MHz}$
is present. This is in contrast to (d) where also multiples of the  
frequencies $\pm A/2$  appear. 
\label{fig:tf}}
\end{figure}

The irradiations at the different frequencies can be clearly observed 
in Fig.~\ref{fig:tf} where the normalized amplitudes of the short-time Fourier 
transform \cite{AM04} are plotted for the relevant cases 
(using the method and implementation of \cite{KHG14}). The important difference
between the controls of Fig.~\ref{fig:vx1} for the polarization transfer from 
$S_z$ to $I_z$ and the controls of Fig.~\ref{fig:vx2} for the transfer from 
$S_z$ to $I_x$ manifest itself in the short-time Fourier transforms of the $v$ 
parts of the controls which are visible in subfigures (c) and (d) of
Fig.~\ref{fig:tf}. In subfigure (c), one notices the characteristic frequency of
$A/2=5 \text{ MHz}$.  In contrast to subfigure (c), 
many more frequencies appear in  
(d) at multiples of the frequencies $\pm A/2$.
This agrees with the square-wave form of the control $v_x$ in Fig.~\ref{fig:vx2}.

In general context, our results can also 
interpreted as  a connection between energy and time
optimizations. The energy optimization leads to a sinusoidal solution
while the time optimization leads to a square wave (see Fig.~\ref{fig:vx2}).
This phenomenon  has been also observed for the simultaneous inversion of two uncoupled
spins  \cite{ALZBGS10,AAPPTZGS12} where the minimum energy solution
was related to the first harmonic in the Fourier expansion of the time-optimal
solution. The same reasoning applies here to the solutions of Section~\ref{sec:theory2}.

\section{Non-sinusoidal carrier wave forms \label{gen}}
We discuss now the  possibility 
and opportunities of non-sinusoidal carrier wave forms for the  electron and nuclear spin.
Here, we focus on the nuclear spin as the maximal amplitude $\vmax$ limits the minimum polarization transfer times
in the rotating frame. 
But similar arguments might also be used for the micro-wave carrier wave form in applications where the
minimum duration of an experiment is limited by $\umax$.
As explained in Section~\ref{sec:theory}, 
we have considered so far a sinusoidal carrier wave form 
$\tilde{v}_x(t)=
v(t) \cos[\omega_I^{\mathrm{rf}} t + \phi(t)]$,
which is motivated by the limited bandwidth of typical
radio-frequency wave form generators and amplifiers.
Equation~\eqref{opt_pulse} states the time-optimal  radio-frequency Hamiltonian for 
transferring polarization from the state $S_z$ to $I_x$ in the rotating frame. In the lab frame, the corresponding 
radio-frequency Hamiltonian is given by
\begin{equation} \label{nonideal}
H^{\mathrm{rf}}=4\pi \vmax\, \cos(\omega_I^{\mathrm{rf}} t -\pi/2)\,   \mathrm{sgn}[ \sin(\pi A t)]  \, I_x.
\end{equation}
However, it is conceivable (e.g., for applications at low magnetic fields or for nuclei with small gyromagnetic ratios)
that the resonance frequency and the corresponding carrier frequency $\omega_I^{\mathrm{rf}}$ of the controls 
are sufficiently small such that non-sinusoidal wave forms (containing higher harmonics of 
the carrier frequency $\omega_I^{\mathrm{rf}}$) can be created and amplified. 
One can therefore envision a radio-frequency Hamiltonian for the ideal case of infinite bandwidth
as given by
\begin{equation} \label{ideal}
\tilde{H}^{\mathrm{rf}}=4\pi \vmax\, \mathrm{sgn}[\cos(\omega_I^{\mathrm{rf}} t -\pi/2)]\,   \mathrm{sgn}[ \sin(\pi A t)]  \, I_x.
\end{equation}
Assuming the same maximal radio-frequency $2 \vmax$ (in the lab frame)
and switching from the Hamiltonian $H^{\mathrm{rf}}$ in Eq.~\eqref{nonideal}
to the Hamiltonian $\tilde{H}^{\mathrm{rf}}$ in Eq.~\eqref{ideal},
the radio-frequency amplitude of the carrier frequency is implicitly 
increased in the rotating frame by another factor of $4/\pi$ (similarly as in discussed in Section~\ref{sec:disc}).
Consequently, the polarization transfer from the state $S_z$ to $I_x$ would be achievable
using only $(\pi/4)^2 \approx 61.7 \%$ 
of the conventional transfer time.

\section{Conclusion \label{conc}}

We have presented time-optimal polarization transfers from an electron spin to a
nuclear spin for the case of secular hyperfine couplings. In particular, we 
have analyzed the
transfers from the electron-spin state $S_z$ to the nuclear-spin states $I_z$ and
$I_x$. For the transfer to $I_x$, we could improve on the duration of
on-resonance sinusoidal solutions by applying a control which has the form of 
a square wave. Our results also highlight differences
between optimizations for minimum energy and minimum time. We have also 
discussed how these differences are related to bandwidth limitations.

\begin{acknowledgments}
H.Y. acknowledges the financial support from the \emph{Research Grants Council} (RGC) 
of Hong Kong (Grant 538213).
R.Z. and S.J.G. acknowledge support from the \emph{Deutsche Forschungsgemeinschaft} 
(DFG) through grants GL 203/7-1 and GL 203/7-2.
\end{acknowledgments}

\appendix

\section{Decomposition of unitaries\label{app_A}}

\subsection{Unitaries which transfer $2 S_z I_z$ to $I_z$\label{appendix_Iz}}

All unitaries in $\SU(4)$ can be be decomposed as $K_1 A K_2$ with 
a slow evolution $A:=\exp[-i (\alpha S^{\alpha} I_y + \beta S^{\beta} I_y)]$
and fast unitaries $K_1$ and $K_2$  which can be generated by controls 
on the electron spin and the secular hyperfine coupling (cf.\ 
\cite{Khaneja01b,KGB02,ZYK:2008,Helgason00}); recall that 
$S^{\alpha}=\left(
\begin{smallmatrix}
1 & 0 \\
0 & 0
\end{smallmatrix}
\right)
\otimes\sigma_{0}$ and $S^{\beta}=\left(
\begin{smallmatrix}
0 & 0 \\
0 & 1
\end{smallmatrix}
\right)
\otimes\sigma_{0}$.
The unitaries that transfer $2S_zI_z$ to $I_z$ can be determined as solutions to the matrix 
equation $I_z = K_1 A K_2
(2S_z I_z) K_2^{\dagger} A^{\dagger} K_1^{\dagger}$. The fast unitaries
$K_1$ and $K_2$ can be parameterized using canonical coordinates of the second kind
(see Section~2.8 of  \cite{Elliott09}, Section~2.10 of \cite{Var:1984}, or Chapter~III, Section~4.3
of \cite{Bou_Lie:1989}), 
i.e.\
\begin{subequations}\label{eq:K2}
\begin{align}
K_1 := & \; e^{-i a_7 S_x} e^{-i a_6 S_y}  e^{-i a_5 S_z}  \nonumber \\
&  \times e^{-i a_4 2 S_x I_z} e^{-i a_3 2 S_y I_z} e^{-i a_2 2 S_z I_z}  
e^{-i a_1 I_z}, \label{A1a} \\
K_2 := & \; e^{-i b_1 S_x} e^{-i b_2 S_y}  
e^{-i b_3 2 S_x I_z} e^{-i b_4 2 S_y I_z}  \nonumber \\
& \times e^{-i b_5 S_z}  e^{-i b_6 2 S_z I_z}  e^{-i b_7 I_z}. \label{A1b}
\end{align}
\end{subequations}
The surjectivity of the representations in Eq.~\eqref{eq:K2} is verified in Appendix~\ref{app:surj}.
As the unitary $K_1$ commutes with $I_z$ and parts of $K_2$ commute with $2S_z I_z$, 
the matrix equation simplifies to 
\begin{align*}
I_z = &\; A  e^{-i b_1 S_x}  e^{-i b_2 S_y}  e^{-i b_3 2 S_x I_z} 
e^{-i b_4 2 S_y I_z}  (2S_z I_z) \\
& \times e^{i b_4 2 S_y I_z} e^{i b_3 2 S_x I_z} e^{i b_2 S_y} e^{i b_1 S_x}     
 A^{\dagger}.
\end{align*}
With the help of the computer algebra system Maple \cite{maple18}, one can verify that 
either $\alpha = 2 \pi z_1$ and $\beta = \pi + 2 \pi z_2$
or  $\alpha = \pi + 2 \pi z_1$ and $\beta = 2 \pi z_2$ with $z_1,z_2 \in \Z$ holds. 
In Sec.~\ref{sec:theory} of the main text, we focused
on the first case assuming that $\alpha=0$ and $\beta=\pi$  (i.e.\ $z_1=z_2=0$), all  other cases are similar.

\subsection{Unitaries which transfer $2 S_z I_z$ to $I_x$\label{appendix_Ix}}

Similarly as in Appendix~\ref{appendix_Iz}, the unitaries
can be decomposed into 
a product $K_1 A K_2$ of
fast unitaries $K_1$, $K_2$
and a slow evolution
$A=\exp[-i (\alpha S^{\alpha} I_y + \beta S^{\beta} I_y)]$.
In particular, all unitaries which transfer $2 S_z I_z$ to $I_x$ have to satisfy the matrix equation 
$I_x = K_1 A K_2
(2S_z I_z) K_2^{\dagger} A^{\dagger} K_1^{\dagger}$.
By observing trivial commutators, the matrix equation
simplifies to 
\begin{align*}
I_x = &\; e^{-i a_4 2 S_x I_z} e^{-i a_3 2 S_y I_z} e^{-i a_2 2 S_z I_z}  
e^{-i a_1 I_z} A  \\
& \times e^{-i b_1 S_x}  e^{-i b_2 S_y}  e^{-i b_3 2 S_x I_z} 
e^{-i b_4 2 S_y I_z}  (2S_z I_z) \\
& \times e^{i b_4 2 S_y I_z} e^{i b_3 2 S_x I_z} e^{i b_2 S_y} 
e^{i b_1 S_x}  A^{\dagger} \\ 
& \times e^{i a_1 I_z} e^{i a_2 2 S_z I_z} e^{i a_3 2 S_y I_z} e^{i a_4 2 S_x I_z}.
\end{align*}
With the help of the computer algebra system Maple \cite{maple18}, one can infer that 
$\beta = \alpha - \pi + 2 \pi z$ holds for $z \in \Z$. In Sec.~\ref{sec:theory2} of the main text,
we consider the case of $A=\exp[-\pi i S_z I_y ]$ which corresponds to 
$\alpha=\pi/2$, $\beta=-\pi/2$, and $z=0$. This choice is actually 
optimal: It follows from Eq.~\eqref{Eq_complex} in the main text that
\begin{align*}
\alpha=&\int_0^T -2\pi [v_x(t)\sin({\pi At})-v_y(t)\cos(\pi At)]\, dt,\\
\beta=&\int_0^T 2\pi [v_x(t)\sin({\pi At})+v_y(t)\cos(\pi At)]\, dt
\end{align*}
holds for any given time $T$.
Consequently, $\beta-\alpha=\int_0^T 4\pi v_x(t)\sin({\pi At}) dt$. One applies 
the condition $\beta = \alpha - \pi + 2 \pi z$ and obtains
$\int_0^T 4\pi v_x(t)\sin({\pi At}) dt=- \pi + 2 \pi z$. This implies that
\begin{equation*}
|\int_0^T 4\pi v_x(t)\sin({\pi At}) dt|\geq |- \pi + 2 \pi z|\geq \pi.
\end{equation*}
On the other hand, one has
$|\int_0^T 4\pi v_x(t)\sin({\pi At}) dt|\leq \int_0^T 4\pi 
v_{\max}|\sin({\pi At})|dt = 8v_{\max}T$ (see Sec.~\ref{sec:theory2}).
In order to satisfy 
the condition $\beta = \alpha - \pi + 2 \pi z$, the time $T$ has 
to fulfill the inequality $8v_{\max}T \geq \pi$. One gets a lower bound 
$T_{\min}\geq \pi/(8v_{\max})$
on the minimum time $T_{\min}$. In summary, the scheme presented in 
Sec.~\ref{sec:theory2} of 
the main text is optimal as it saturates the lower bound.

\section{Verification of the surjectivity of the representations in Eq.~\eqref{eq:K2} \label{app:surj}} 
In order to verify the surjectivity of $K_1$ in Eq.~\eqref{A1a}
it is sufficient to verify the surjectivity of the product 
\begin{align*}
&\tilde{K}_1 = \tilde{K}_1(a_7,a_6,a_5,a_4,a_3,a_2) :=  \\
&e^{-i a_7 S_x} e^{-i a_6 S_y}  e^{-i a_5 S_z} 
e^{-i a_4 2 S_x I_z} e^{-i a_3 2 S_y I_z} e^{-i a_2 2 S_z I_z}   \\
\end{align*}
which consists of the first six elements of $K_1$ as the seventh element
commutes with all the other ones. First we show that there exists 
$a_4'$, $a_3'$,  and $a_2'$ such that
\begin{align}
 e^{-i a_7 S_x} e^{-i a_6 S_y}  e^{-i a_5 S_z}e^{-i a_4 2 S_x I_z} e^{-i a_3 2 S_y I_z} e^{-i a_2 2 S_z I_z}= \nonumber\\
e^{-i a_4' 2 S_x I_z} e^{-i a_3' 2 S_y I_z} e^{-i a_2' 2 S_z I_z}e^{-i a_7 S_x} e^{-i a_6 S_y}  e^{-i a_5 S_z},
\label{eq:comm}
\end{align}
which can be written as 
\begin{align*}
&e^{-i a_4' 2 S_x I_z} e^{-i a_3' 2 S_y I_z} e^{-i a_2' 2 S_z I_z} =  
e^{-i a_7 S_x} e^{-i a_6 S_y}  e^{-i a_5 S_z} \\
& \times (e^{-i a_4 2 S_x I_z} e^{-i a_3 2 S_y I_z} e^{-i a_2 2 S_z I_z})
e^{i a_5 S_z}  e^{i a_6 S_y} e^{i a_7 S_x}.
\end{align*}
The effect of the conjugation with $\exp(-i a_5 S_z)$ is 
\begin{align*}
&e^{-i a_5 S_z}e^{-i a_4 2 S_x I_z} e^{-i a_3 2 S_y I_z} e^{-i a_2 2 S_z I_z}e^{i a_5 S_z}\\
&=e^{-i a_5 S_z}e^{-i a_4 2 S_x I_z}e^{i a_5 S_z} \\
&\times e^{-i a_5 S_z}e^{-i a_3 2 S_y I_z}e^{i a_5 S_z}e^{-i a_5 S_z} e^{-i a_2 2 S_z I_z}e^{i a_5 S_z}\\
&=e^{-ia_4[\cos(a_5)2S_xI_z+\sin(a_5)2S_yI_z]}\\
&\times e^{-ia_3[\cos(a_5)2S_yI_z-\sin(a_5)2S_xI_z]}e^{-i a_2 2 S_z I_z}\\
&=e^{-i \a_4'' 2 S_x I_z} e^{-i a_3'' 2 S_y I_z} e^{-i a_2'' 2 S_z I_z},
\end{align*}
where the last step follows from the Euler-angle decomposition. Similar arguments for
the conjugations with $e^{-i a_6 S_y}$ and $e^{-i a_7 S_x}$ demonstrate Eq.~\eqref{eq:comm}.
Any element in the connected Lie group that is infinitesimally generated by the elements $-iS_x$, $-iS_y$, $-iS_z$,
$-i2 S_x I_z$, $-i 2 S_y I_z$, and $-i2 S_z I_z$ can be achieved by a finite product of elements having the form of
$\tilde{K}_1$; this is a consequence of Lemma~6.2 in \cite{JS72}. We apply
Eq.~\eqref{eq:comm} and the Euler-angle decomposition multiple times and obtain 
$\tilde{K}_1(a_7,a_6,a_5,a_4,a_3,a_2) \tilde{K}_1(\tilde{a}_7,\tilde{a}_6,\tilde{a}_5,\tilde{a}_4,\tilde{a}_3,\tilde{a}_2)
=\tilde{K}_1(c_7,c_6,c_5,c_4,c_3,c_2)$ 
for certain values of $c_7$, $c_6$, $c_5$, $c_4$, $c_3$, and $c_2$.
In summary, we have verified the surjectivity of the representations $\tilde{K}_1$ and
$K_1$.

Similar as for Eq.~\eqref{eq:comm}, one can verify
that 
\begin{align*}
&e^{-i a_5 S_z} e^{-i a_4 2 S_x I_z} e^{-i a_3 2 S_y I_z}\\
&= e^{-i a_4' 2 S_x I_z} e^{-i a_3' 2 S_y I_z} e^{-i a_5 S_z}
\end{align*}
holds for some $a_4'$ and $a_3'$. Consequently, the surjectivity of $K_1$ implies the
surjectivity of $K_2$.

An alternative second argument for the surjectivity of Eq.~\eqref{A1a} applies the decomposition $K'_1 A' K'_2$
for the set $K= \exp(\fk)$  of all fast operations where $K'_i=\exp(\fk')$ and $A'=\exp(\fa')$. This decomposition
is a consequence of the Cartan decomposition $\fk = \fk' \oplus \fp'$ where the corresponding linear subspaces are given by
$\fk':=\mathrm{span}\{ -i S_x, -i S_y, -i S_z, -i I_z \}$, $\fp':=\mathrm{span}\{-i 2 S_x I_z, -i 2 S_y I_z, -i 2 S_z I_z\}$,
and the abelian subalgebra $\fa':=\mathrm{span}\{-i 2 S_z I_z\}\subseteq \fp'$
\cite{Helgason00}. The decomposition $K'_1 A' K'_2$ implies that the decomposition
\begin{align*}
U'  = U e^{i \pi S_z I_z} = &\; e^{-i d_1 S_x} e^{-i d_2 S_y} e^{-i d_3 S_z} e^{-i d_4 2 S_z I_z} \\
& \times e^{-i d_5 S_z} e^{-i  d_6 S_x}
e^{-i d_7 S_y} e^{-i d_8 I_z}
\end{align*}
is a surjective parameterization of the set of all fast operations. Therefore, the surjectivity is also 
verified for 
\begin{align*}
U = &\; e^{-i d_1 S_x} e^{-i d_2 S_y} e^{-i d_3 S_z} e^{-i d_4 2 S_z I_z} \\
       & \times e^{-i d_5 S_z} e^{-i d_6 S_x} e^{-i d_7 S_y} e^{-i d_8 I_z} e^{-i \pi S_z I_z}\\
    = &\; e^{-i d_1 S_x} e^{-i d_2 S_y} e^{-i (d_3+d_5) S_z} e^{-i (d_4 + \pi/2) 2 S_z I_z} \\
       & \times e^{i  d_6 2 S_y I_z} e^{-i  d_7 2 S_x I_z}  e^{-i d_8 I_z} \\
    = &\; e^{-i d_1 S_x} e^{-i d_2 S_y} e^{-i d_3' S_z} e^{-i d_4'  2 S_x I_z} \\
       & \times e^{-i  d_5' 2 S_y I_z} e^{-i  d_5' 2 S_z I_z}  e^{-i d_8 I_z},
\end{align*}
where the last equality follows from the Euler-angle decomposition. This completes the second argument
for the surjectivity of Eq.~\eqref{A1a}.

%\bibliography{epr}


\end{document}